# Grid Stability and Power Factor Dynamics in Solar Farms Integration

Hassan A. Osseily [#1], Omar M. Srouji [*2], Mohammad I. Al Jammal [#3]

[#1] Electrical & Electronics Department, Lebanese International University & The International University of Beirut
Beirut - Lebanon

[1] `hassan.osseily@liu.edu.lb`
[2] `11830576@students.liu.edu.lb`
[3] `11630124@students.liu.edu.lb`

*Abstract—* This paper examines the impact of solar farm fluctuations on grid stability, focusing on maintaining an optimal power factor. ETAP-based simulations and case studies are used to analyze real-time grid performance under solar variability. Reactive power control strategies and advanced inverter functions are proposed for stabilization. Theoretical analysis and simulation results highlight effective integration techniques. Artificial intelligence is trailed for controlling the SVC in adaptive reactive power compensation. The study provides practical solutions for improving reliability in renewable-integrated power systems.

## I. INTRODUCTION

In mid-20th century, solar power began with photovoltaic cells; in 1954 [1] silicon-based solar cells with 6% efficiency were invented. Over decades, solar cell design improved efficiency, reduced cost, and enabled worldwide use. Solar farm grid connection became common in the 1990s [2]. Researchers developed control strategies for issues like voltage instability caused by mismatched solar production and demand, such as in microgrids [3]. One strategy was energy storage and demand response systems [4]. Recently, power electronics and machine learning combined in smart grids enabled efficient controls like real-time monitoring and adaptive control, reducing fluctuations and ensuring stability [5].Studies show grid-connected solar farms cause instability, especially as solar capacity grows.

### A. *Solar Farm Integration Stability:*

Arnett and Patel [6] examined the impact of integrating solar PV on grid stability. They studied how variable solar generation affects frequency and voltage, causing instability.

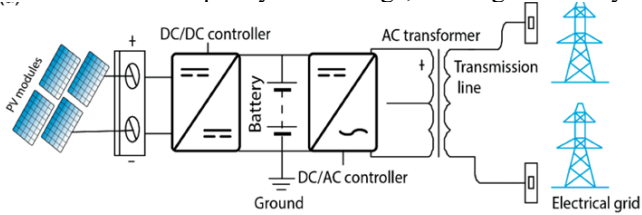

Fig. 1. Component of Grid-Connected Solar Farm

To mitigate this, they assessed control methods like frequency support and voltage regulation. The study provides a theoretical framework but lacks practical validation. The tactics were tested in simulations and do not consider real conditions like severe weather or grid overloads. Fig.1 shows components of the grid-connected solar farm.

### B. *Reactive Power Management in Solar Farms:*

Muthukumar and Khadkikar [7] focused on reactive power control for grid-connected solar PV. Their study described conventional techniques for voltage stability control, including advanced inverter controls and reactive power compensators. These strategies reduced voltage swings and improved stability in grids with varying solar penetration. Fig.2 shows the reactive power capability of PV versus the reactive power requirement.

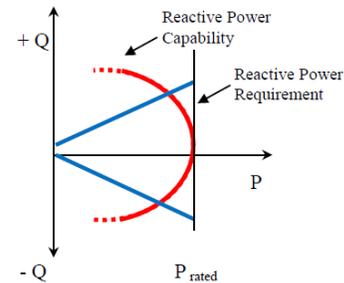

Fig. 2. Reactive Power Requirement Versus Reactive Power Capability of PV.

## II. PROBLEM STATEMENT AND PROPOSED SYSTEM

Although efficient, these techniques rely on static control mechanisms [8], which may not suit the dynamic nature of modern grids with growing solar penetration. The study does not consider real-time predictive technology. Research should explore AI and ML for reactive power management [9], as these adaptive methods can better anticipate and respond to voltage instability. This study examines how solar farms dynamically affect grid performance, focusing on power factor. Using ETAP simulations [10], it analyzes solar power fluctuations on stability and evaluates control methods like reactive power management and advanced inverters. The combined theory and simulations highlight strategies to reduce solar variability and improve reliability. Fig.3 shows the system block diagram. As a case study, this paper analyzes the impact of Abu Dhabi's solar farm (1 GW, largest worldwide) on utility grid stability.

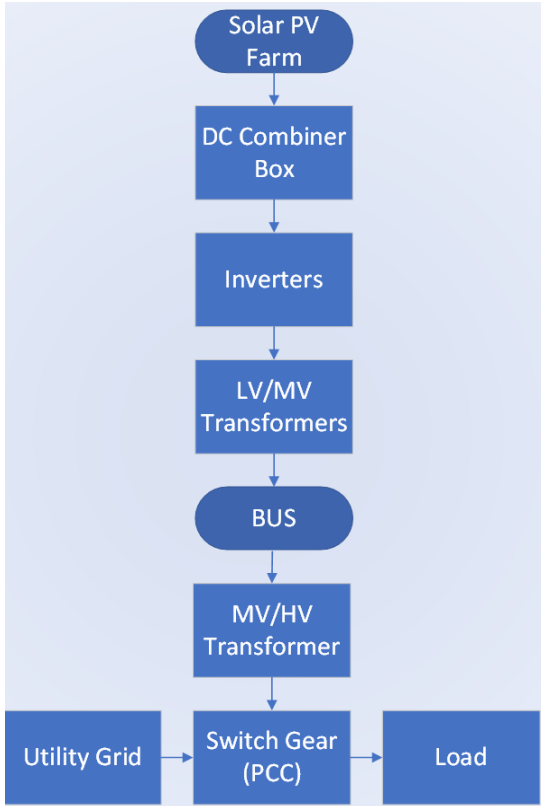

Fig. 3. Block Diagram of the System

The block diagram shows the connection to the grid and switchgear at the point of common coupling (PCC) to feed the load also show grid-tied inverter which is the main component.

A. *Grid Tied Inverters:*

The Grid-Tied Inverter (GTI) in Fig.4 is a high-capacity inverter for utility-scale PV systems, the SMA Sunny Central SC 4200 UP-US. It converts DC power from PV arrays into AC power sent directly to the grid. This inverter is essential for large solar projects, ensuring reliable grid connection and efficient energy conversion.

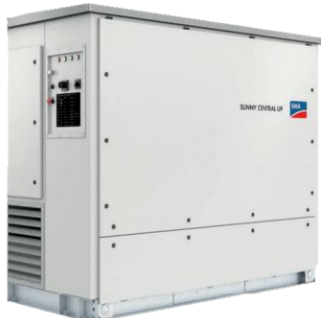

Fig. 4. Grid-Tied Inverter

Utility-scale systems that are built solely for grid-connected photovoltaic (PV) generation without the addition of energy storage solutions are best suited for this arrangement. Table. I shows the specification of the used inverter.

TABLE I
SMA INVERTER SPECIFICATIONS

| Specification | SC 4200 UP |
|---|---|
| Max. DC Voltage | 1500 V |
| MPP Voltage Range (25°C / 50°C) | 921–1325 V / 1050 V |
| Min. / Start DC Voltage | 891 V / 1071 V |
| Max. DC Current | 4750 A |
| Max. Short-Circuit Current | 8400 A |
| Number of DC Inputs | 24 double pole / 32 single pole fu |
| Nominal AC Power (35°C / 50°C) | 4200 kVA / 3780 kVA |
| Nominal AC Active Power (cos φ = 0.8) | 3360 kW / 3024 kW |
| Nominal AC Current (35°C / 50°C) | 3850 A / 3465 A |
| Nominal AC Voltage / Range | 630 V / 504–756 V |
| Frequency / Range | 50 Hz (47–53 Hz) / 60 Hz (57–63 |
| Max. Efficiency / Euro / CEC Efficiency | 98.8% / 98.7% / 98.5% |
| Cooling System | OptiCool (intelligent air coolin |
| Degree of Protection (Electronics) | IP54 |
| Dimensions (W × H × D) | 2815 × 2318 × 1588 mm |
| Weight | < 3700 kg |
| Operating Temperature Range | –25°C to +60°C |
| Self-Consumption (max / standby) | < 8100 W / < 370 W |
| Communication Interfaces | Ethernet, Modbus Master/Slav |

B. *System's Hierarchical Structure*

The structure is shown in the Single Line Diagram (SLD) in Fig. 5. The PV field has 8,400 panels rated at 500 W each, arranged in 25-module series strings with about 1,139 V string voltage at max power. There are 336 such strings in parallel, forming a PV array delivering 4.2 MW peak DC power. DC power is converted to 630 V AC using high-efficiency grid-tied inverters rated in MW. The figure shows a repeated part of the system.

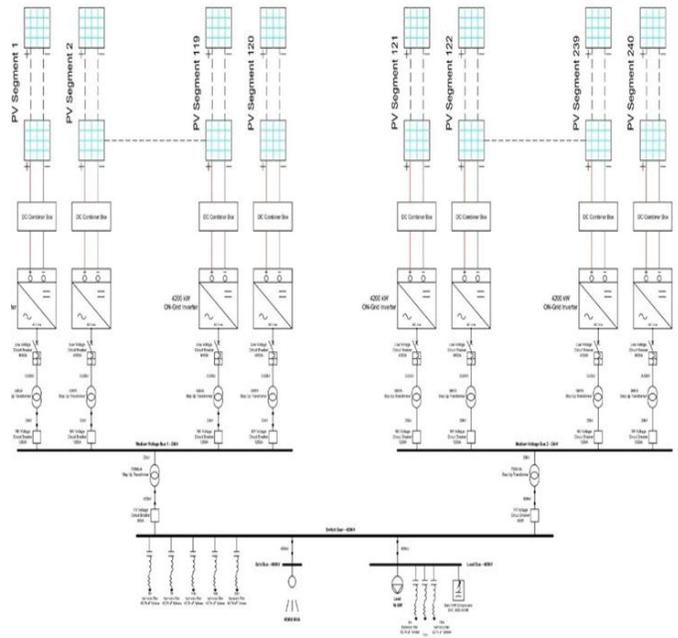

Fig. 5. Block Diagram of the System

The AC output is stepped up from 0.63 kV to 33 kV through 240 MV transformers (5 MVA each) aggregating PV outputs. These feeders lead to the main substation, where two HV transformers (750 MVA each) raise voltage to 400 kV for Abu Dhabi's grid. The transmission system is modeled as a 40 GW swing bus representing a strong grid for large generation absorption. Power is dispatched via switchgear to an 18 GW load on a 400 kV overhead line. SVCs and harmonic filters at key nodes stabilize voltage and reduce THD, aiding grid compliance and stability. Table II summarizes components and quantities.

TABLE II
LIST OF ALL NEEDED COMPONENTS

| Component | Specification | Count | Capacity/Feature |
|---|---|---|---|
| PV Panels | LONGi Hi-MO 5 (500 W per panel) | ~2,016,000 | 1 GW total capacity |
| Inverter | SMA Sunny Central 4200 UP | 240 | 4.2 MW per inverter, grid-tied operation |
| Transformers | LV/MV Step-up transformers (0.63 kV to 33 kV) | 240 | 5 MVA per unit |
| Transformers | MV/HV Step-up transformers (33 kV to 400 kV) | 2 | 750 MVA per unit |
| Monitoring Systems | 1 per sub-block | 240 | Tracks voltage, current, and energy output |
| Reactive Power Compensation | STATCOM (Siemens SVC PLUS) | 1 | ±6500 MVAR per unit |
| Central Control System | Plant-wide | 1 | Real-time monitoring and control |
| Protection Relays | Overcurrent relays | 100 | 1 per sub-block |
| Harmonic Filters | 42.74 µF | 8 | 3 Load bus / 5 Switch Gear |
| Breakers | LV Circuit breakers | 240 | 1 per Sub-block |
| Breakers | MV Circuit breakers | 240 | 1 per Sub-block |
| Breakers | HV Circuit breakers | 2 | 1 per block |
| AI (Artificial Intelligence) | SVC Control | - | - |

### III. FLOW CHART OF THE SYSTEM

Fig.6 shows the simulation steps: maintain power factor above 0.95, add SVC if below, rerun load flow, then assess harmonics using THD.

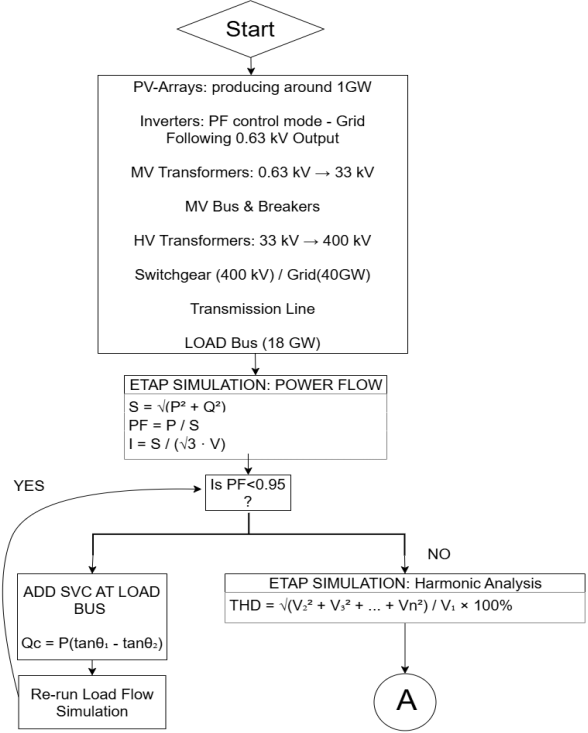

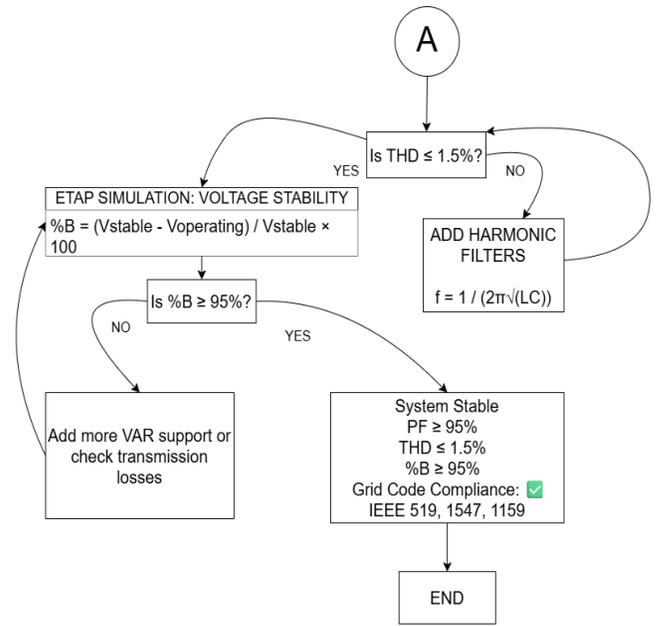

Fig. 6. Flowchart of the System

### IV. DESIGN STEPS AND CALCULATION

This section presents the electrical design and analysis of a 1 GW PV plant integrated into a 40 GW grid serving an 18 GW load. It models PV arrays, inverters, transformers, and components using ETAP. Key criteria include grid code compliance, reactive power support, harmonic mitigation, and voltage stability. The plant uses 8400 panels in 336 parallel strings (25 panels each). Each array outputs 4.2 MW DC, converted to 630 V AC via inverters, stepped to 33 kV by 240 MV transformers (5 MVA), then to 400 kV by two HV transformers (750 MVA). Power feeds the 18 GW load on a 400 kV line via switchgear. SVCs and filters control voltage and harmonics. Required calculations include:

1) *Apparent Power:*
$$S = \sqrt{(P^2 + Q^2)} \quad (1)$$

2) *Power Factor (PF):*
$$PF = P / S \quad (2)$$

3) *Line Current:*
$$I = S / (\sqrt{3} \times V) \quad (3)$$

4) *Required Compensation:*
$$Q_c = P \times (\tan\theta_1 - \tan\theta_2) \quad (4)$$

5) *Total Harmonic Distortion (THD):*
$$THD = \sqrt{(V_2^2 + V_3^2 + ... + V_n^2)} / V_1 \times 100\% \quad (5)$$

6) *Filter Tuning Frequency:*
$$f = 1 / (2\pi \sqrt{(LC)}) \quad (6)$$

7) *Voltage Stability Margin:*
$$\%B = (V_{stable} - V_{operating}) / V_{stable} \times 100 \quad (7)$$

8) Total Panels:
$$N_{total} = N_s \times N_p = 25 \times 336 = 8400 \quad (8)$$

9) Array Output:
$$P_{array} = N_{total} \times P_{panel} = 8400 \times 500W = 4.2 \text{ MW} \quad (9)$$

10) Required Arrays:
$$N_{arrays} = 1000 \text{ MW} / 4.2 \text{ MW} \approx 238 \quad (10)$$

11) Inverter AC Output:
$$P_{inverter} = P_{DC} / ILR = 4.2 / 1.2 = 3.5 \text{ MW} \quad (11)$$

12) Inverter Current:
$$I_{inv} = P_{array} / (\sqrt{3} \times V_{inv} \times PF) \quad (12)$$

13) LV Transformer Current:
$$I_{LV} = 5 \times 10^6 / (\sqrt{3} \times 630) \approx 4584 \text{ A} \quad (13)$$

14) MV Transformer Current:
$$I_{MV} = 5 \times 10^6 / (\sqrt{3} \times 33000) \approx 87.5 \text{ A} \quad (14)$$

15) HV Transformer S:
$$S = 500 / 0.98 \approx 510.2 \text{ MVA} \quad (15)$$

16) HV Current:
$$I \approx 736 \text{ A} \quad (16)$$

17) Recommended Rating:
$$S \times 1.3 \approx 663.3 \text{ MVA} \quad (17)$$

18) SVC Reactive Power:
$$Q_c = 17{,}774.68 \times 0.3673 \approx 6500 \text{ MVAR} \quad (18)$$

19) Filter Capacitance per Phase:
$$C = Q\phi / (2\pi f V_{LN}^2) \quad (19)$$

These formulas calculate total power drawn, power efficiency, and current in a three-phase system, including SVC compensation and reactive power. This section establishes a full-scale, code-compliant design for a 1 GW PV system. Key components are modeled and sized with ETAP, and all power flow, transformer, and compensation calculations are verified. Voltage regulation, harmonic filtering, and power factor correction ensure safe and efficient grid integration. This foundation supports advanced control and optimization studies in later chapters.

## V. RESULTS USING ETAP SIMULATION

### A. Power Factor Improvement

Power factor (PF) measures how efficiently electrical power is converted into useful work. A high PF means most power drawn from the grid is effectively used, reducing transmission losses, improving voltage regulation, and maximizing transformer and equipment capacity. Conversely, a low PF causes higher currents for the same real power, leading to increased line losses ($I^2R$), possible transformer and cable overloads, and voltage drops. This reduces system efficiency, raises operating costs, and may incur utility penalties. After applying SVCs and harmonic filters, the system's power factor significantly improved, reaching near unity and enhancing overall efficiency as shown in Fig.7.

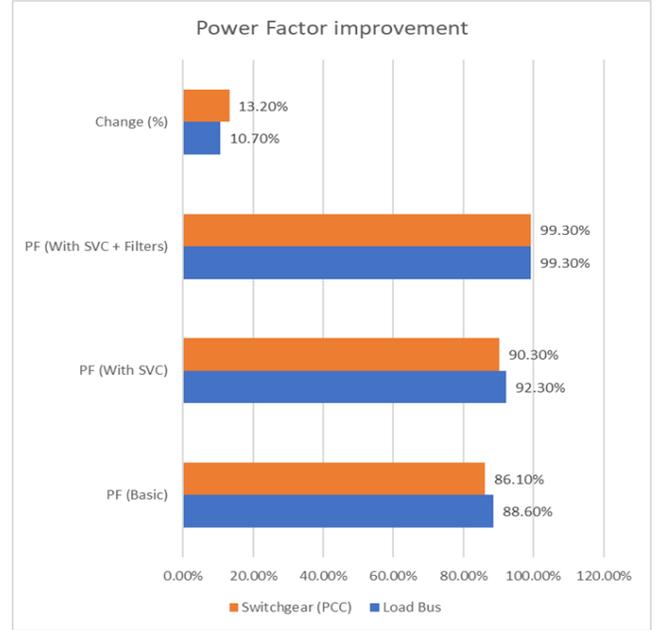

Fig. 7. Power Factor Improvement

### B. Voltage Stability and %B

The %B value (Voltage Reserve Margin) measures how close a power system is to voltage collapse by indicating extra load capacity before instability (Fig.8). Higher %B reflects a more stable system. %B above 95% is healthy, offering good voltage support, while below 90% suggests proximity to instability and blackout risk.

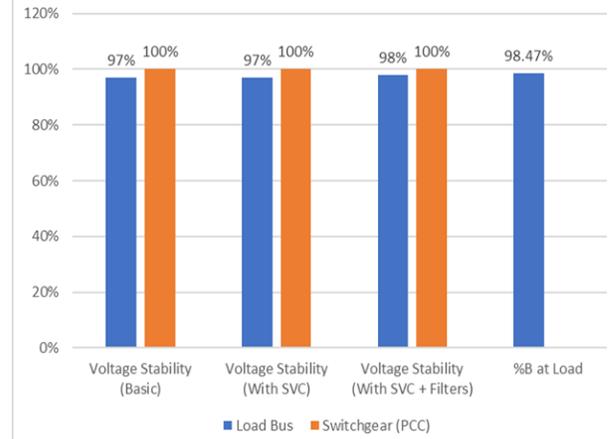

Fig. 8. Voltage Stability and %B

### C. Harmonic Distortion (THD)

Harmonic filters greatly reduced total harmonic distortion, improving power quality. THD dropped from 19.48% to 1.32% at the switchgear and from 9.03% to 0.40% at the load bus, within IEEE 519 limits (Fig.9).

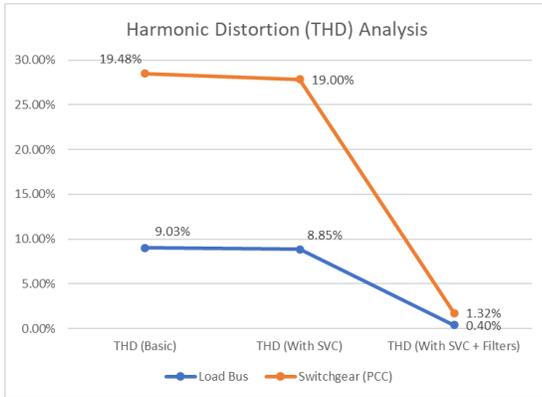

Fig. 9. Harmonic Distortion (THD) Analysis by ETAP

### D. Power Flow Comparison (MW and MVAR)

Compensation reduced MVAR demand at both load bus and switchgear, improving transmission efficiency (Fig.10). The switchgear maintained positive MW export, reflecting stable system real power flow.

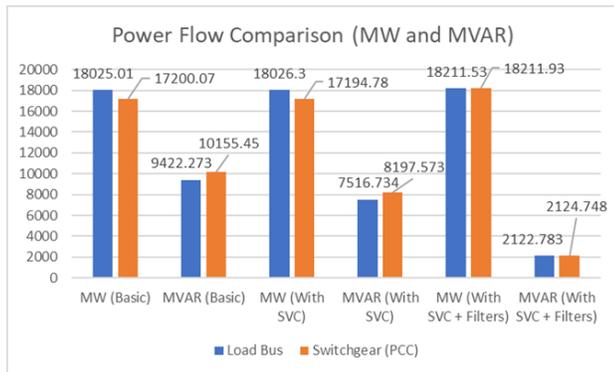

Fig. 10. Power Flow Comparison (MW and MVAR)

### E. Overcompensation and Power Factor Anomalies at PCC

Load flow analysis showed 99.3% power factor and efficient transfer at the load bus, but overcompensation at the switchgear with distorted power factor (-90.6%) due to excess reactive power from fixed SVC and filters (Fig.11). This may cause inefficiencies and relay miscoordination. AI-based adaptive SVC control is proposed for dynamic reactive power optimization.

| ID | MW | Mvar | Amp | %PF |
|---|---|---|---|---|
| Switch Gear | -18211.930 | -2124.748 | 26566.2 | 99.3 |
| LOADBUS | 18210.110 | 2124.740 | 26563.6 | 99.3 |
| Switch Gear | 17542.620 | -8207.332 | 27954.7 | -90.6 |

Fig. 11. Load Flow by ETAP

### F. AI-Based SVC Controller for Voltage Regulation

A reinforcement learning controller using Deep Q-Network (DQN) was developed to regulate SVC reactive power injection in real time. The controller maintains voltage levels within IEEE limits (0.95–1.05 pu). During training, the DQN agent interacts with a simulated system, learning actions by minimizing voltage deviations. The system state (e.g., voltage) is input to the neural network, which outputs control decisions. Based on rewards, the agent updates its policy over episodes. Once trained, the DQN adjusts reactive power to stabilize voltage. Fig. 12 shows the AI controller's response to fluctuations.

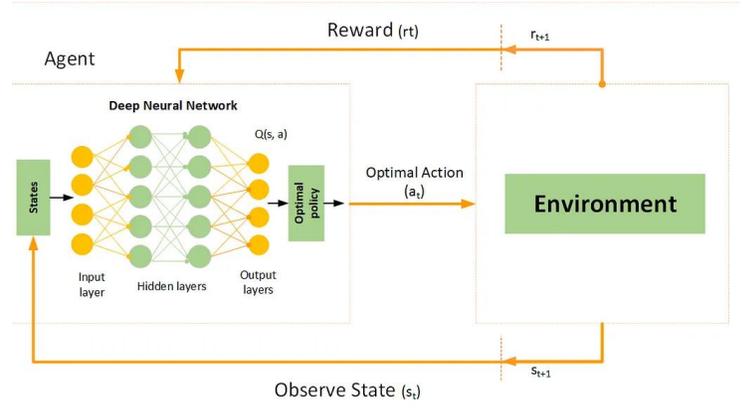

Fig. 12. AI Logic

### G. Overall Interpretation

The AI-based SVC controller effectively corrects voltage disturbances within one control cycle, acting only as needed and maintaining voltage near nominal values. It requires no precise system model, adapts to noise and nonlinearities, and can be expanded for enhanced control. Future improvements include online learning and integration of additional power quality metrics (Fig.13 and Fig.14).

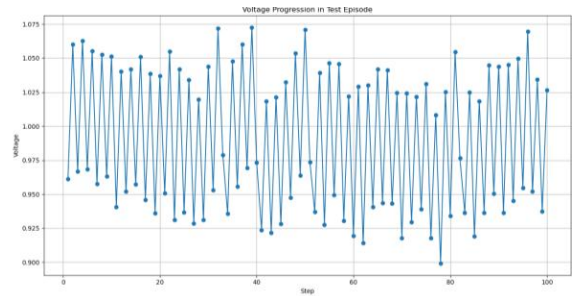

Fig. 13. Voltage Progression in a Test Episode

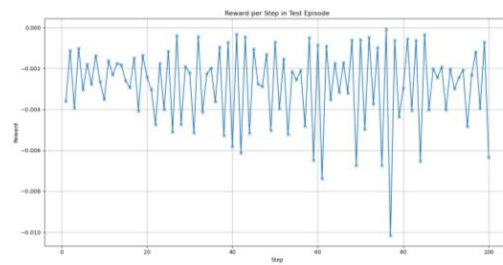

Fig. 14. Reward Per Step in the Same Episode

## H. Compliance with IEEE Standards

The system was evaluated against IEEE 519-2014 (harmonics) and IEEE 1547-2018 (interconnection and power factor). Key results: THD below 1.5%, power factor above 95%, voltage stability and %B margins above 95%, ensuring compliance and reliable grid integration (Table III).

TABLE III
COMPLIANCE SUMMARY TABLE

| Parameter | IEEE Standard Requirement | Achieved Value in Final System |
|---|---|---|
| Voltage THD | ≤ 1.5% (for >161 kV grids) | 0.40% (Load Bus), 1.32% (Switchgear) |
| Power Factor (PF) | ≥ 95% leading/lagging | 99.3% at both Load Bus and Switchgear |
| Voltage Stability (% Index) | ≥ 95% (industry standard) | 98% (Load Bus) |
| Voltage %B (Reserve Margin) | Ideally ≥ 95% | 98.47% (Load Bus) |

## I. Comparison of Reactive Power Compensation Strategies

The basic system showed low power factor, moderate voltage stability, and high harmonic distortion. Adding an SVC improved power factor and voltage stability but failed to reduce harmonic distortion adequately (Fig.15).

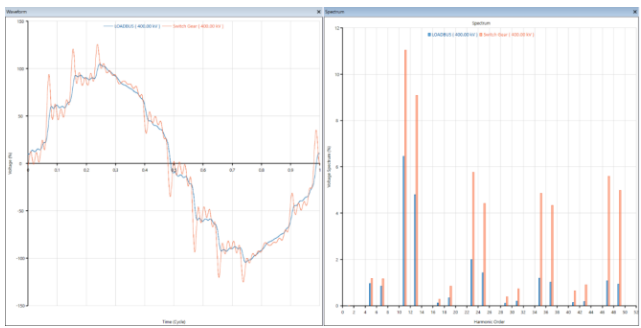

Fig. 15. Voltage Waveform & Individual Harmonics Spectrum Before Compensation

Combining SVCs with tuned harmonic filters yielded the best results, raising %B to 98.47%, enhancing voltage stability, reducing harmonic distortion, and lowering reactive power demand (Fig.16). This combination is crucial for large-scale solar farm grid integration.

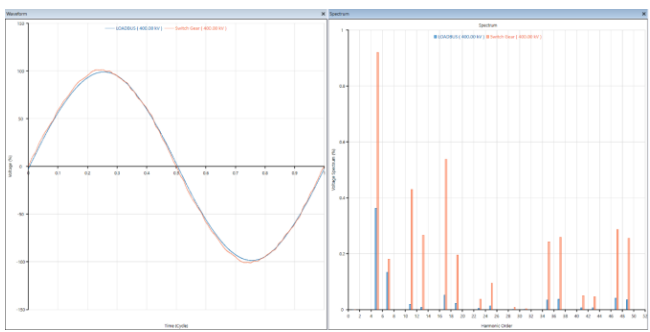

Fig. 16. Voltage & Individual Harmonics Spectrum After Compensation

## VI. CONCLUSION

This study analyzed a 1-GW solar farm integration focusing on reactive power, voltage stability, harmonics, and system performance. SVCs improved voltage support, reduced reactive power, and raised power factor but couldn't reduce harmonics beyond IEEE limits. Adding single-tuned harmonic filters suppressed harmonics, improving power quality and compliance. Combined SVCs and filters achieved 99.3% power factor, 98% voltage stability, 98.47% reserve margin, 1.32% THD at switchgear, and 0.40% at load bus. Real-world examples like Noor Abu Dhabi and Benban Solar Park confirm these strategies, reflecting global requirements for stable, compliant solar integration.

## REFERENCES

[1] R. Baker, "The role of solar energy in power systems: A global overview," Renew. Sustain. Energy Rev., vol. 125, p. 109800, 2021.

[2] A. Jäger-Waldau, "PV power to reach 1,000 GW by 2020: An overview of the current status of photovoltaics," Renew. Sustain. Energy Rev., vol. 78, pp. 169–180, 2019.

[3] H. A. Osseily, "Design and Analysis of a Micro-grid for a Lebanese Village," Rare Metal Materials and Engineering, vol. 53, no. 6, pp. –, Jul. 2024.

[4] C. Zhang and L. Xu, "Improved power factor control in power systems with high penetration of solar photovoltaic generation," IEEE Trans. Power Syst., vol. 35, no. 3, pp. 1912–1921, 2020.

[5] H. Liu and Y. Zhang, "The evolution and future of grid integration for photovoltaic systems," Energy, vol. 235, p. 121357, 2021.

[6] R. Arnett and A. Patel, "Impact of solar PV on grid stability and control mechanisms," Energy Reports, vol. 6, pp. 765–772, 2020.

[7] S. Muthukumar and V. Khadkikar, "Reactive power control in grid-connected solar photovoltaic systems," Renew. Energy, vol. 143, pp. 902–910, 2020.

[8] S. Yang and X. Zhao, "Advanced inverter control for solar energy systems: A review," J. Renew. Sustain. Energy, vol. 11, no. 5, p. 050301, 2019.

[9] H. Zhang, Y. Wang, and Z. Li, "Artificial intelligence-based predictive control for energy management in smart grids," IEEE Transactions on Smart Grid, vol. 12, no. 3, pp. 2050–2060, May 2021. doi: 10.1109/TSG.2020.3025890.

[10] M. A. Khan, S. U. Rehman, and F. A. Khan, "AI-Based Optimal Load Flow Analysis of Grid-Connected PV Systems Using ETAP," IEEE Access, vol. 10, pp. 56321–56330, 2022. doi: 10.1109/ACCESS.2022.3173124.